\documentclass[reprint,amsmath,amssymb,aps,prc]{revtex4-2}
\usepackage{graphicx} 
\usepackage{dcolumn} 
\usepackage{bm} 
\usepackage{siunitx}
\usepackage{float}
\usepackage{caption}
\usepackage{subcaption}
\usepackage{hyperref}  
\hypersetup{colorlinks = true ,linkcolor={blue},citecolor={red},urlcolor={blue}}  
\begin{document}
\title{A polynomial in transverse momentum manifesting thermalisation in nuclear collisions}
\author{Rahul Ramachandran Nair}
\email{physicsmailofrahulnair@gmail.com}
\thanks{\\Orcid ID: 0000-0001-8326-9846}
\affiliation{National Centre For Nuclear Research, 02-093 Warsaw, Poland}
\altaffiliation{The work was performed while the author was affiliated with the National Centre For Nuclear Research, Warsaw, Poland.\\Present address: Sreeraj Bhavan, Karukachal PO, Kottayam(District), Kerala, India- 686540} 
\date{\today}
\begin{abstract}
A specific value of light front variable $\zeta_c(p_T,m)$ of the inclusively produced hadrons in a relativistic nuclear collision is constructed as a polynomial in transverse momentum $p_T$ of the particle within the UrQMD transport model. It is shown that those particles with the absolute value of their light front variable $\zeta$ greater than the corresponding $\zeta_c(p_T,m)$ follow a Fermi-Dirac or Bose-Einstein distribution depending upon the spin of the particle.
\end{abstract}
\maketitle
\begin{figure*}[ht!]
\centering
\includegraphics[width=0.95\textwidth]{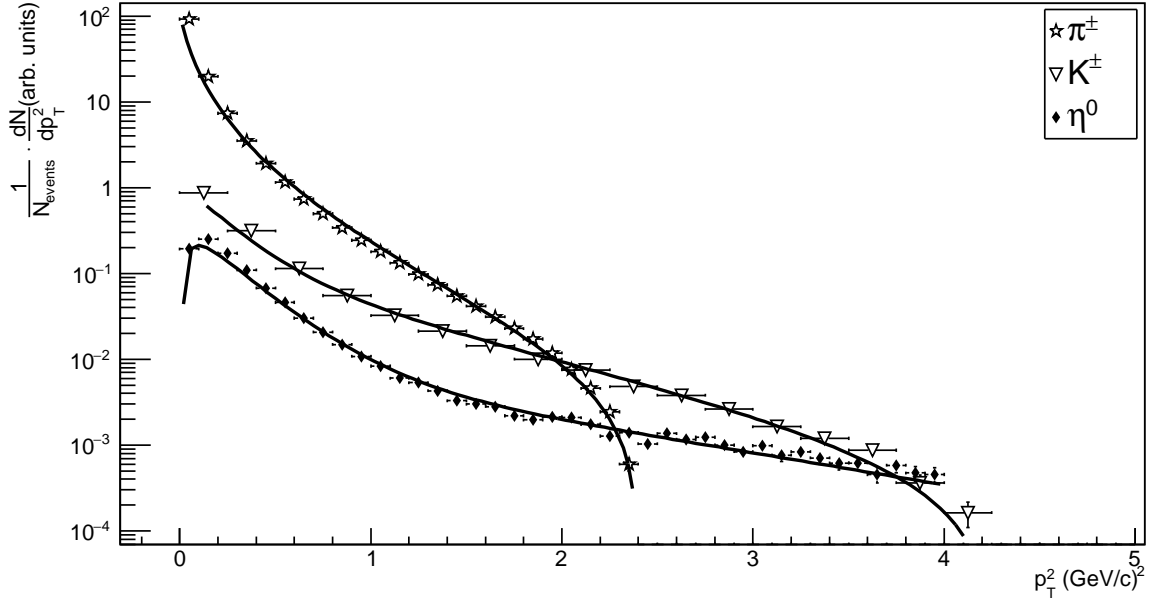}
\caption{$p_T^2$ distributions of mesons with $|\zeta^{\pm}| > \zeta_{c}(p_T,m)$ fitted with Eq.\eqref{PtSqInt}. Solid red curves are the result of the fits.}
\label{ptsqmesons}
\end{figure*}
\begin{figure*}[ht!]
\centering
\includegraphics[width=0.95\textwidth]{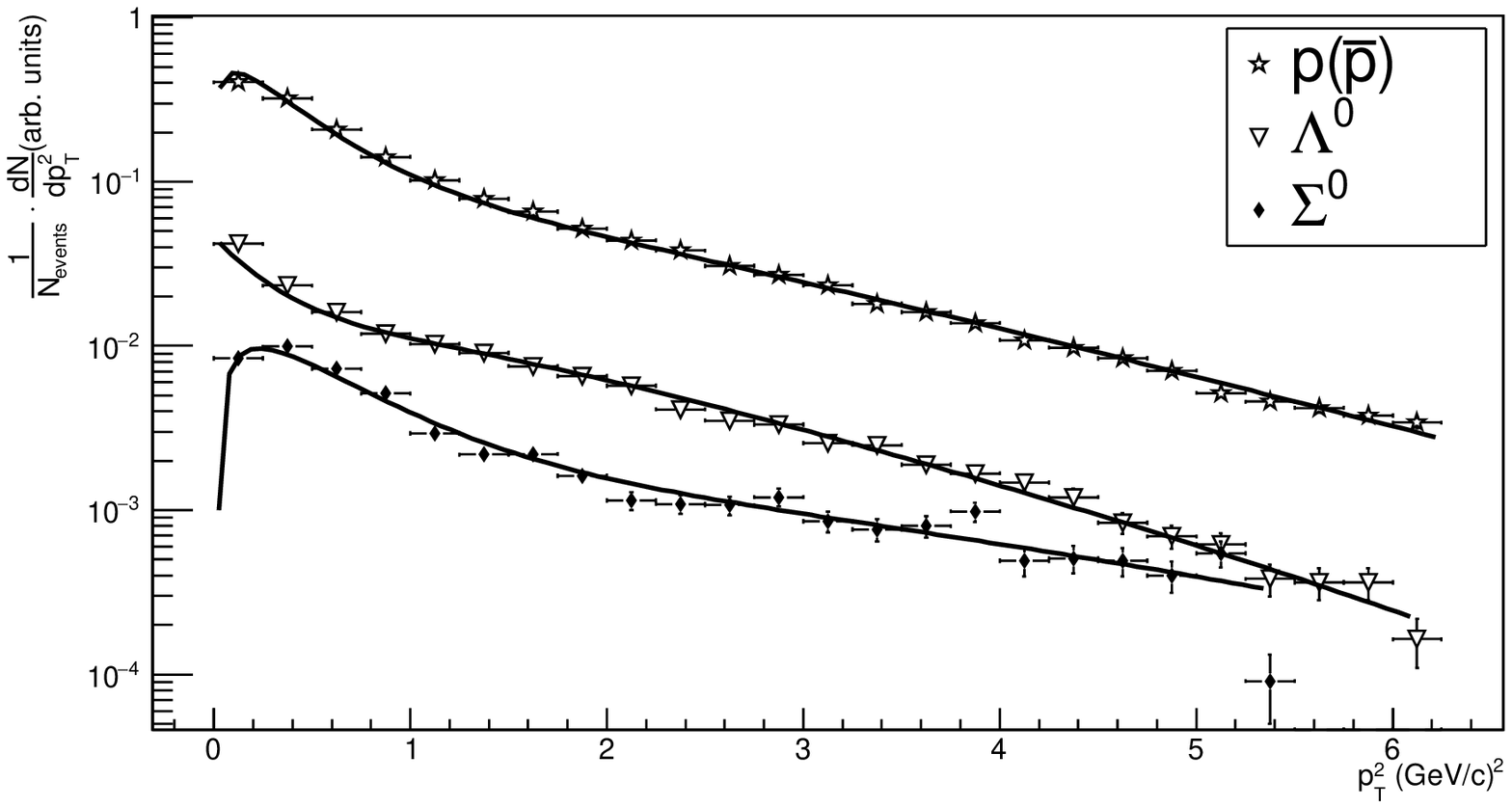}
\caption{$p_T^2$ distributions of baryons with $|\zeta^{\pm}| > \zeta_{c}(p_T,m)$ fitted with Eq.\eqref{PtSqInt}. Solid curves are the result of the fits.}
\label{ptsqbaryons}
\end{figure*}

The light front variables introduced by Dirac \cite{Dirac49} can be written for the particles produced in a hadron-hadron or nucleus-nucleus collision in the following scale and Lorentz invariant form:
\begin{equation} \label{EqnXi}
\xi^{\pm} = \pm \frac{E + |p_{Z}|}{\sqrt{s}} \text{ and }
\zeta^{\pm} = \mp \ln\left[\xi^{\pm}\right]
\end{equation}
where, $s$ is the Mandelstam variable, $p_{Z}$ is the z-component of the momentum and $E$ is the energy of the particle with the positive and negative signs corresponding to the two hemispheres \cite{Garsevanishvili78, Garsevanishvili79,  Amaglobeli99, Djobava03, Chkhaidze2006}. A constant value of $\zeta$ corresponds to a paraboloid in the phase space of the particles. It was observed in the earlier studies with the experimental data that a certain constant value of the light front variable denoted by $\tilde\zeta$ can be used to select a group of particles that supposedly has reached thermal equilibrium \cite{Garsevanishvili78, Garsevanishvili79,  Amaglobeli99, Djobava03, Chkhaidze2006}. In the work presented here, we show that the transverse momentum ($p_T$) dependence of the $\tilde\zeta$ can be quantified to select a group of thermalised particles in a pseudorapidity region of $|\eta| < 1 $ in the phase space.\\\\
Fifty five thousand minimum bias Au-Au collisions are simulated at $\sqrt{s} = 200$ GeV using the Ultrarelativistic Quantum Molecular Dynamics (UrQMD-version 3.4) model of heavy ion collisions for this purpose \cite{BASS1998255,Bleicher_1999}. The $\zeta$ distributions are made for $\pi^{\pm}$ , $K^{\pm}$, $\eta^{0}$, $p(\bar{p})$, $\Lambda^{0}$ and  $\Sigma^{0}$
with $|\eta| < 1 $ in different $p_T$ intervals. Taking in to account the constraints on $\eta$ and $p_T$, these distributions are fitted with the following equation
\begin{equation}
\frac{dN}{d\zeta} \sim \int_0^{p_{T,max}^2} E f(E)dp_T^2
\label{ZetaInt}
\end{equation}
while $f(E)$ has the Fermi-Dirac form ($[e^{E/T}+1]^{-1}$) for the baryons and Bose-Einstein form ($[e^{E/T}-1]^{-1}$) for the mesons. The upper limit of the integral in Eq.\eqref{ZetaInt} is given by
\begin{equation}
p_{T,max}^2 = (\xi\sqrt{s})^{2} - m^{2} 
\end{equation}
The lowest value of $\zeta$ down to which the fit can be successfully performed is taken as $\tilde{\zeta}$ for the $p_T$ range with which the distribution is made for. If the $p_T^2$ and polar angle distributions of the particles with $\zeta > \tilde{\zeta}$ can also be described with the respective form of $f(E)$, then $\tilde{\zeta}$ is assigned as the constant $\zeta_c$ for the $p_T$ range. If the $\tilde{\zeta}$ does not follow this condition, then a larger value of $\zeta$ is considered as the new $\tilde{\zeta}$ and the procedure is repeated until a $\zeta_c$ is found for the range of $p_T$ under consideration. A detailed analysis of the similar kind within a single $p_T$ range with a cut on $\eta$ can be seen in \cite{nair2021feasibility}. The procedure is repeated in several $p_T$ ranges for each of the species. From this analysis, an approximate polynomial relationship of the form 
\begin{equation}\label{polynomial}
\zeta_{c}(p_T,m) =  a_{0}+ a_{1}p_T + a_{2}p_T^2 + a_{3}p_T^3
\end{equation}
between $\zeta_{c}$ and $p_T$ is constructed. Thus for a specific particle, we have a $\zeta$ from Eq.\eqref{EqnXi} and a $\zeta_{c}(p_T,m)$. from Eq.\eqref{polynomial}. In the next step, we make the $p_T^2$ distributions of those particles with their $|\zeta^{\pm}| > \zeta_{c}(p_T,m)$. The resulting distributions are found to be describable with the following equation
\begin{equation}
\frac{dN}{dp_T^2} \sim \int_0^{p_{z,max}} f(E)dp_z
\label{PtSqInt}
\end{equation}
taking into account the kinematic constraints from $p_T$ and $\eta$ cuts. The upper limit of the integral in Eq.\eqref{PtSqInt} is given by 
\begin{equation}
p_{z,max}  = \frac{m^2 + p_{T}^2 - (\sqrt{s}e^{ -\zeta_{c}(p_T,m)})^2}{-2\sqrt{s}e^{ -\zeta_{c}(p_T,m)}}
\label{pzmax}
\end{equation}
Note that the $\zeta_{c}(p_T,m)$ in Eq.\eqref{pzmax} are the polynomials in Eq.\eqref{polynomial} with the coefficients as given in TABLE \ref{TableIntegral} for various species. The temperatures obtained from the fit for various species of hadrons are reported in TABLE.\ref{TableIntegral}. The results of the fit for the $p_T^2$ distributions of mesons are shown in FIG. \ref{ptsqmesons} and those for the baryons are shown in FIG. \ref{ptsqbaryons}. The uncertainties in the distributions are statistical. ROOT (version- 6) software package \cite{BRUN199781} is used for making kinematical distributions from the simulated data, performing the integration and fitting.
\begin{table*}[hbt!]
\begin{center}
\setlength{\tabcolsep}{10pt} 
\renewcommand{\arraystretch}{1.5} 
\begin{tabular} {llllllll}
\hline\hline \noalign{\smallskip}
Species       &$a_{0}$& $a_{1}$& $a_{2}$ & $a_{3}$    & T(MeV) &$\chi^2/n.d.f$  \\
\noalign{\smallskip}\hline\noalign{\smallskip}
$\pi^{\pm}$   & \num{6.55804e+0} & -\num{1.89233e+0}  & \num{6.27206e-01}  & -\num{7.39660e-02}  & 170$\pm$4  & 10/22 \\
$K^{\pm}$     & \num{6.04416e+0} & -\num{9.38750e-01}  & -\num{1.54280e-02} & \num{5.75811e-02}  & 199$\pm$6  & 4/14 \\
$\eta^{0}$    & \num{5.99366e+0} & -\num{9.72565e-01}  & \num{1.32247e-01}  & -\num{1.01038e-02} & 185$\pm$2  & 40/38 \\
$p(\bar{p})$  & \num{5.36285e+0} & -\num{2.56643e-01}  & -\num{2.15872e-01} & \num{5.08619e-02}  & 232$\pm$3  & 12/23  \\
$\Lambda^{0}$ & \num{5.14367e+0} & \num{1.59066e-02}   & -\num{4.08923e-01} &  \num{9.85914e-02} & 219$\pm$4  & 10/22 \\
$\Sigma^{0}$  & \num{5.16500e+0} & -\num{3.37435e-01}  & \num{4.65282e-03 } & -\num{2.56765e-02} & 233$\pm$5  & 20/19 \\
\noalign{\smallskip}\hline\hline
\end{tabular}
\end{center}
\caption{Results of the fit of $p_T^2$ distributions of particles with their $|\zeta^{\pm}| > \zeta_{c}(p_T,m)$ using the relation in Eq.\eqref{PtSqInt}}
\label{TableIntegral}
\end{table*}
\\\\It is therefore concluded that within the context of the UrQMD microscopic transport model of relativistic heavy-ion collisions, we can establish a polynomial in $p_T$ for a certain constant value of the light front variable which can select a thermalised group of inclusively produced hadrons. This effect establishes two paraboloids for a specific particle. One corresponding to the $\zeta$ value of the particle and the second corresponding to the $\zeta_{c}(p_T,m)$ of the particle. Thus a grouping of thermalised particles is made depending upon the relative position of the two paraboloids in the phase space for a given particle. In this sense, the existence of $\zeta_{c}(p_T,m)$ in the phase space is analogous to the general theory of relativity where the observed dynamics is governed by geometrical entities. It also suggests that there exist a non-trivial integral relation between the transverse and longitudinal components of the momentum of these hadrons. The measured temperatures for the particles are larger than the Hagedorn temperature and lattice QCD predictions of temperatures about the phase transition \cite{Gazdzicki2016, Ratti_2018} which hints towards the possible existence of a deconfined state of quarks and gluons in the medium where the hadrons are emitted from. Hence this kind of analysis may be used to test the presence of a thermalised quark-gluon plasma in the heavy-ion collisions at RHIC and LHC from the maximum entropy principle itself. The exact reason for the origin of this effect in high energy nucleus-nucleus collisions, experimental verification and the extent of its profoundness has to be explored further. Performing the same analysis with experimental data at very high values of transverse momentum, carrying out the HBT interferometry, two-particle correlation and collectivity studies for the particles with $\zeta > \zeta_{c}(p_T,m))$ and $\zeta < \zeta_{c}(p_T,m))$ might be handy in this respect \cite{Heinz1999,Wiedemann99,Heinz2013}. A similar analysis with the hadrons produced in hadron-hadron collisions at LHC might be helpful in better understanding the observed collectivity \cite{doi:10.1146/annurev-nucl-101916-123209,bhalerao2020} in such systems. The conclusions are drawn within the context of a phenomenological model and it is imperative to perform the analysis with the experimental data available at RHIC and LHC.\\\\
\textit{I would like to thank Prof. Marcus Bleicher (UrQMD collaboration) for the consent to use the UrQMD source code. I thank Dr Bharati Naik (IIT, Bombay), Prof. Kajari Mazumdar(TIFR, Mumbai), Dr Preeti Dhankher (University of California, Berkeley) and Dr Souvik Priyam Adhya (Charles University, Prague) for helping me with the computational resources.}
\bibliography{article}
\end{document}